\documentclass[conference]{IEEEtran}
\IEEEoverridecommandlockouts
\usepackage{cite}
\usepackage{amsmath,amssymb,amsfonts}
\usepackage{algorithmic}
\usepackage{graphicx}
\usepackage{textcomp}
\usepackage{xcolor}
\def\BibTeX{{\rm B\kern-.05em{\sc i\kern-.025em b}\kern-.08em
    T\kern-.1667em\lower.7ex\hbox{E}\kern-.125emX}}

\usepackage{comment}
\usepackage{subcaption}
\usepackage{caption}
\usepackage{amsmath}

\usepackage{booktabs} 

\IEEEoverridecommandlockouts                              

\begin{document}

\title{\LARGE \bf
A Design Methodology for Post-Moore’s Law Accelerators: The Case of a Photonic Neuromorphic Processor
}

\author{Armin Mehrabian$^{1}$, Volker J Sorger$^{1}$, and Tarek El-Ghazawi$^{1}$
\thanks{$^{1}$Authors are with the department of Electrical and Computer Engineering of The George Washington University. (email:armin@gwu.edu; sorger@gwu.edu; tarek@gwu.edu) }\\
}
\IEEEoverridecommandlockouts
\IEEEpubid{\makebox[\columnwidth]{\copyright2020 IEEE \hfill} \hspace{\columnsep}\makebox[\columnwidth]{ }}

\maketitle

\pagestyle{empty}

\begin{abstract}
Over the past decade alternative technologies have gained momentum as conventional digital electronics continue to approach their limitations, due to the end of Moore’s Law and Dennard Scaling. At the same time, we are facing new application challenges such as those due to the enormous increase in data. The attention, has therefore, shifted from homogeneous computing to specialized heterogeneous solutions.  As an example, brain-inspired computing has re-emerged as a viable solution for many applications. Such new processors, however, have widened the abstraction gamut from device level to applications. Therefore, efficient abstractions that can provide vertical design-flow tools for such technologies became critical. Photonics in general, and neuromorphic photonics in particular, are among the promising alternatives to electronics. While the arsenal of device level toolbox for photonics, and high-level neural network platforms are rapidly expanding, there has not been much work to bridge this gap. Here, we present a design methodology to mitigate this problem by extending high-level hardware-agnostic neural network design tools with functional and performance models of photonic components. In this paper we detail this tool and methodology by using design examples and associated results. We show that adopting this approach enables designers to efficiently navigate the design space and devise hardware-aware systems with alternative technologies. 
\end{abstract}

\section{Introduction}
 With the rise of artificial intelligence (AI) applications, data-intensive workloads have surged. These, in part result in plateaued speed and energy efficiency of digital von-neumann computers. Many alternative technologies and computing paradigms have been proposed. Photonics is one of these technologies, which has been a major driver of data communication over the past decades. 
 One of the main challenges facing a new technology is the limited and inconsistent availability of design and simulation tools. The field of photonic computing suffers from a wide abstraction gap in design and simulation tools. Most of such tools are currently focused on the device \cite{lumerical2014solutions} and low circuit level \cite{chrostowski2016schematic}. To compete with conventional electronics, there needs to be a long-term effort to devise tools that complete the design flow stack from high-level specification and synthesis to device and technology attachment. Even further, for neuromorphic applications, the stack needs to incorporate top-level functionalities such as those in training and inference of neural networks. Some recent works in photonics have taken this route to bridge the vertical gap by developing application-specific photonic software stacks \cite{anderson2020roc}\cite{bangari2019digital}.\par
 Here, we propose a design methodology applicable to neuromorphic systems. Our methodology is based on extending existing commonly used neural network packages, such as Google Tensorflow. We propose to extend the hardware-agnostic arithmetic units with functional and measurement models of the technology, here photonics. 
 Our approach is distinguished from other similar works in three major ways. First, our approach allows users to benefit and rely on low-level and mid-level features of Tensorflow such as high-speed back-end processing on a variety of hardware choices such as CPUs and GPUs. Secondly, our work particularly emphasizes on \textit{noise} as a significant component of any analog circuit including photonics. Lastly, familiarity with a widely-used platform such as Tensorflow, shortens the learning time and the time to import existing work into our tool.

\section{Design Methodology}
As discussed in the previous section, we propose to extend Tensorflow with models of actual photonic components commonly used in photonic neuromorphics. Our goal is two fold, first, to investigate the effect of non-ideal analog photonic components on the functional performance of a neural network. Secondly, estimate the power consumption of these analog photonic components in such networks to give us a better understanding of the trade-offs of adopting the neuromorphic photonics. In the rest of this section we first introduce a few of the most commonly used photonic components in neuromorphic photonics. Then, we briefly discuss the overall hierarchy of the Tensorflow tool and where and how it was extended. Lastly, we provide example mathematical descriptions of the modeled components.

\subsection{Photonic Components} The recent increased popularity of photonics is mainly due to its low operating power and high bandwidth \cite{miller2017attojoule}. Recently, a multitude of neuromorphic photonic processors have been proposed and even realized \cite{nahmias2018neuromorphic}\cite{bangari2019digital}\cite{mehrabian2018pcnna}\cite{mehrabian2019winograd}\cite{shen2017deep}. In these architectures, basic arithmetic operations are realized by photonic devices that mimic those functionalities. Table \ref{tab:tensorflow_mapping_1} lists some of these arithmetic operations and their corresponding photonic realizations.\par 
 We extend Tensorflow with two classes of models. First, functional models, that transform ideal noise-free arithmetic operations with their realistic analog photonic representations. Secondly, power models that aim to compute power estimates. While power models do not affect the functional performance of a neural network such as the prediction accuracy, functional models influence them.\par
Here we start by introducing a set of commonly adopted photonic devices. We emphasize on two example devices used to realize photonic multiplication, namely micro-ring resonator (MRR) and Mach-Zehnder interferometer (MZI). The two devices realize the same functionality so we use them as an example case for design space exploration.\par
\begin{figure}[ht]
    \centering
    \includegraphics[width=0.9\linewidth]{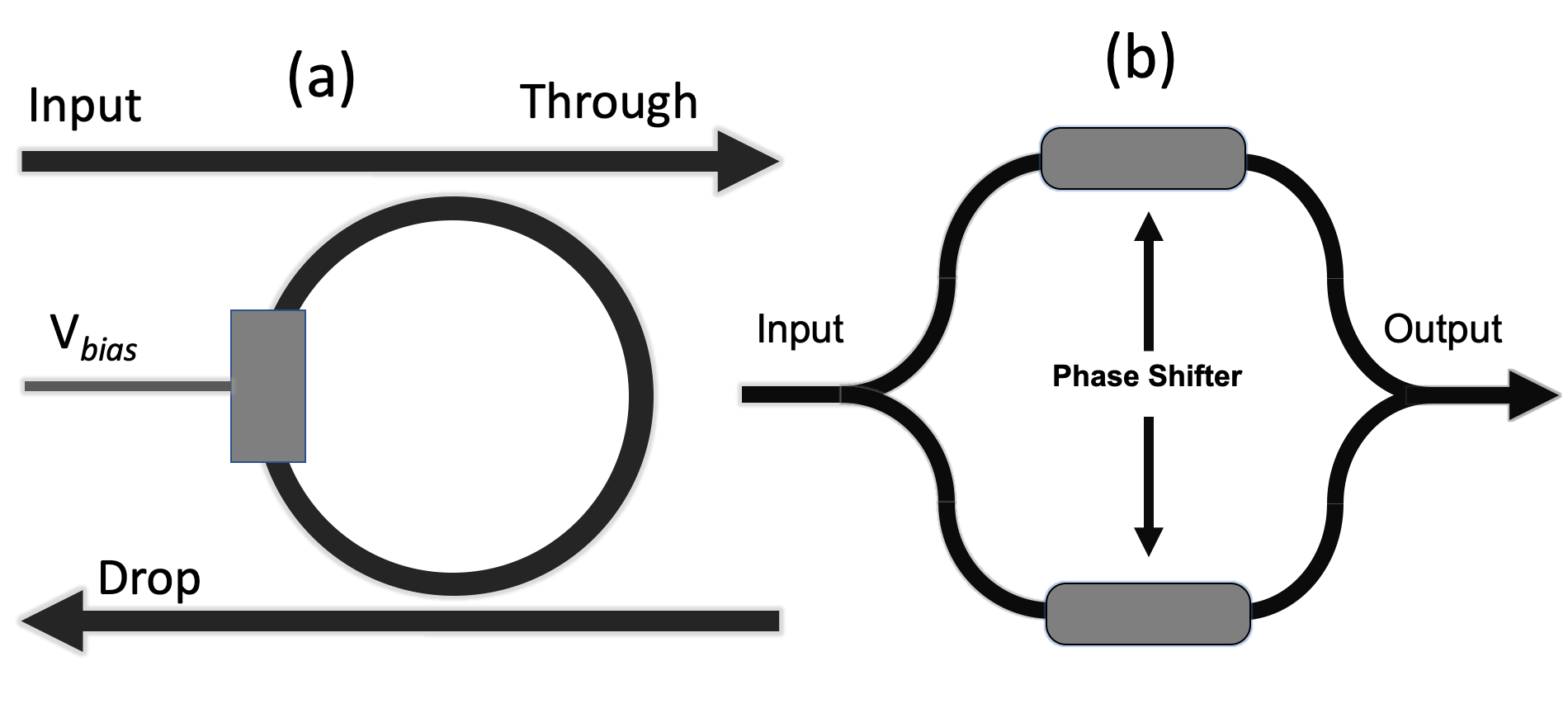}
    \caption{Schematic diagram of (a) a MRR device (b) a MZI device. } 
    \label{fig:mrr_mzi}
\end{figure}

MRRs play a significant role in photonics. A generic MRR is a circular optical waveguide as shown in Figure \ref{fig:mrr_mzi} (a). The MRR in Figure \ref{fig:mrr_mzi} is coupled to one \textit{Through} and one \textit{Drop} waveguides. The portion of the light coupled to the ring will loop through the ring and then couple back to the \textit{Through} waveguide and create anywhere between a destructive or a constructive interference. The level of interference depends on the wavelength of the incoming beam and the resonant frequency of the ring. By applying a bias voltage $V_{bias}$ the resonant frequency of the ring is changed, thus affecting the level of interference. 

That being said, the two outputs of the MRR together can be used to create differential weighting between an incoming light beam and a bias voltage. It should be note that this weighting is spectrally sensitive and even can be engineered to realize selective parallel multiplications on different wavelengths.

Another alternative device that can be used for weighting in photonics is the Mach-Zehnder Interferometer (MZI) \cite{shen2017deep}. Figure  \ref{fig:mrr_mzi} (b) demonstrates a MZI device. The input light beam is split into two beams through a beamsplitter. Each beam incurs a different phase change by a phase shifter. At the output, a combiner combines the two phase-shifted beams. The output beam will have a different amplitude dictated by the relative phase of the two beams, which similar to a MRR can cause a range of interferences. Hence, by controlling the amount of phase shift, a weighting mechanism between the input beam and the phase shift is realized.

In photonics, the summation operation can be achieved optically in two main ways; incoherently via a photodiode or alternatively, coherently by combining two phase-stabalized photonic beams. By feeding a set of input light beams to a photodiode, we can add the power of the beams and generate an electrical current proportional to the sum of the incident beams. \par
Another important class of components in neural networks is the nonlinear activation function. Without nonlinear activation functions the whole neural network collapses into a linear transformation, incapable of finding complex nonlinear tasks. There has been many recent works in photonics to build nonlinear activation functions for neural networks \cite{george2018neural}\cite{miscuglio2018all}. One way to build a nonlinear activation function in photonics is to map the nonlinear activation function onto the transfer function of an electro-optic modulator (EOM). The advantage of this method is that when paired with a photodiode, the output of photodiode is an electrical current, which can directly be used to drive an electro-optic modulator without the need of any direct electrical to optical conversion. Furthermore, we can use a new laser source to be modulated by our signal, which allows to keep signal cascadability high. 
\begin{table}[!t]
\renewcommand{\arraystretch}{1.3}
\caption{Mapping of primitive math operations to their hardware realization.}
\label{tab:tensorflow_mapping_1}
\centering
\begin{tabular}{c|c}
\toprule
Math Operation & Photonic Representation  \\
\midrule
 Multiplication & MRR, MZI \\
 Addition & Photodiode \\
 Connection & Waveguide \\
 Non-linear Activation & Electro-Optic Modulator\\
\bottomrule
\end{tabular}
\end{table}
\subsection{Google Tensorflow}\label{sec:tensorflow}
\begin{figure*}[ht]
    \centering
    \includegraphics[width=.91\linewidth]{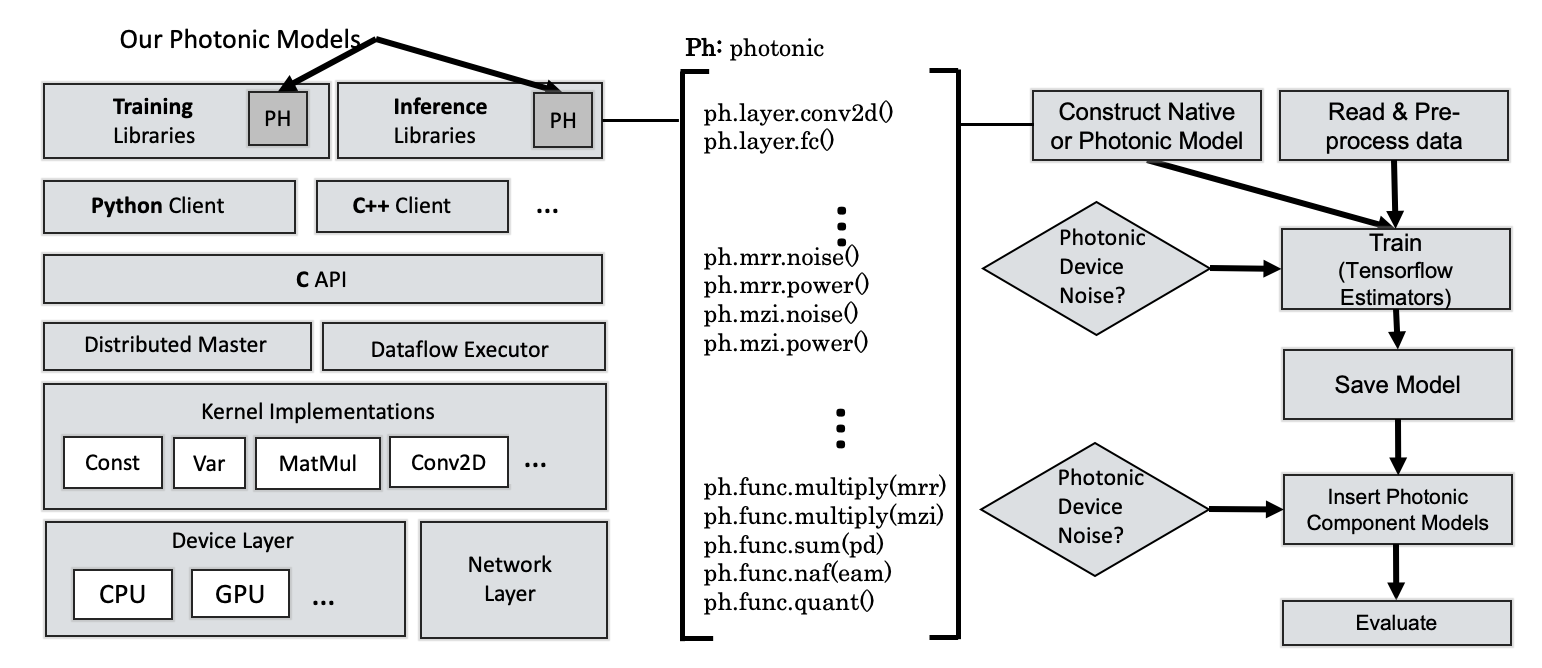}
    \caption{Overview of the Tensorflow architecture and our extended photonic model library implementation.} 
    \label{fig:tf_hierarchy}
\end{figure*}

Tensorflow at heart is a dataflow graph processor that can map a computational graph across machines in a cluster and across different computational devices, such as CPUs, GPUs, and TPUs. While our design methodology is for the most part focused on the inference, the availability of training algorithms allow the designer to benefit from a wide variety of state-of-the-art train-time tools on top of a familiar user interface. Figure \ref{fig:tf_hierarchy} depicts the hierarchical architecture of Tensorflow and our extended photonic models.\par

Core Tensorflow is coded in C++ to take advantage of its performance and portability. Given an input graph, it partitions the graph into sub-graphs to be used by supported underlying computing hardware. From Figure \ref{fig:tf_hierarchy} it can be seen that, many of standard kernels are fused in the low-level kernel implementations to gain better performance for standard neural network architectures. Within the low-level kernel layer, the kernels form a gamut of operations from very simple tensor definition to more complex convolutional and recurrent layers. Since these fused kernels are accessible through high-level Python and C++ clients, we can extend these base kernels inside the training and inference libraries.

\subsection{Extended Models}\label{sec:models}

In the rest of this section, we provide example mathematical models used in this work. First example is the power model for the photodiode. Power in a photodiode is calculated using the \textit{Responsivity} as follows,
\begin{equation}
    R=\frac{I_{ph}}{P_{in}}=\lambda\frac{q}{hc}\eta \;\;\;\; [\dfrac{A}{W}]
\end{equation}
where $P_{in}$ is the power of input incident light, $I_{ph}$ is the photo-current, $q$ is the electron charge, $\lambda$ is the wavelength, $h$ is the Planck’s constant, and $c$ is the speed of light. For a photodiode, given the, technology the \textit{Responsivity} is known. In this work we use values from foundry processes \cite{timurdogan2018aim}.

Another example of a model we implemented here is the noise models. Noise models fall under the functional models class as they perturb the operation of otherwise an ideal photonic neural network. For the same photodiode, there are two types of noise sources namely Thermal noise and the Shot noise, which are derived from,
\begin{equation}
    I_{sn}=\sqrt{2q(I_{ph}+I_D)\Delta f} \quad \textrm{and} \quad I_{tn}=\sqrt{\frac{4K_BT\Delta f}{R_{SH}}}
\end{equation}
where $I_D$ is the dark current, $\Delta f$ is the bandwidth, $K_B$ is the Boltzmann constant, $T$ is temperature in Kelvins and $R_{SH}$ is the total equivalent shunt resistance. Noise models are particularly interesting because they let us explore the design space of photonic neural networks with different noise characteristics.

The last class of models are functional models that aim to create a more realistic implementation of photonic devices or adjust for functional imperfections of photonic hardware. For example in MRRs, which are used to realize the weighting operation, the actual transfer function of the \textit{Through} port is defined by,
\begin{equation}
    T_{Through}=\frac{I_{pass}}{I_{input}}=\frac{r_2^2a^2 - 2r_1r_2acos\phi + r_1^2}{1-2r_1r_2acos\phi+(r_1r_2a)^2}
\end{equation}
where $a$ is the attenuation, $r_1$ and $r_2$ are coupling coefficients with Through and Drop waveguides, and $\phi$ is the single pass phase shift. As a result, when $a$ becomes non-negligible the weighting of the incident beam and the bias voltage incur some level of precision loss.

\section{Results}
\begin{figure*}[ht]
\begin{subfigure}{\textwidth}
  \centering
  \includegraphics[height=0.2\linewidth]{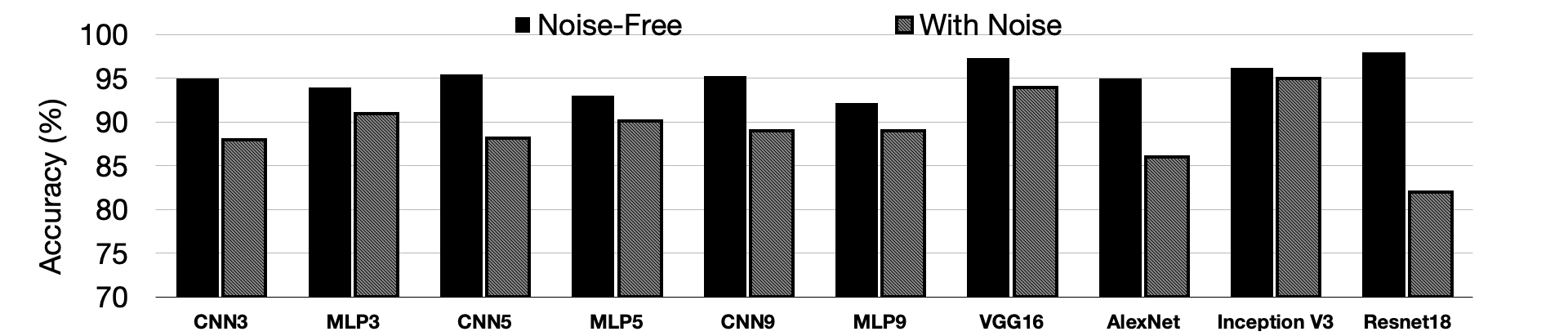}  
  \caption{MRR}
  \label{fig:sub-first}
\end{subfigure}
\begin{subfigure}{\textwidth}
  \centering
  \includegraphics[height=0.2\linewidth]{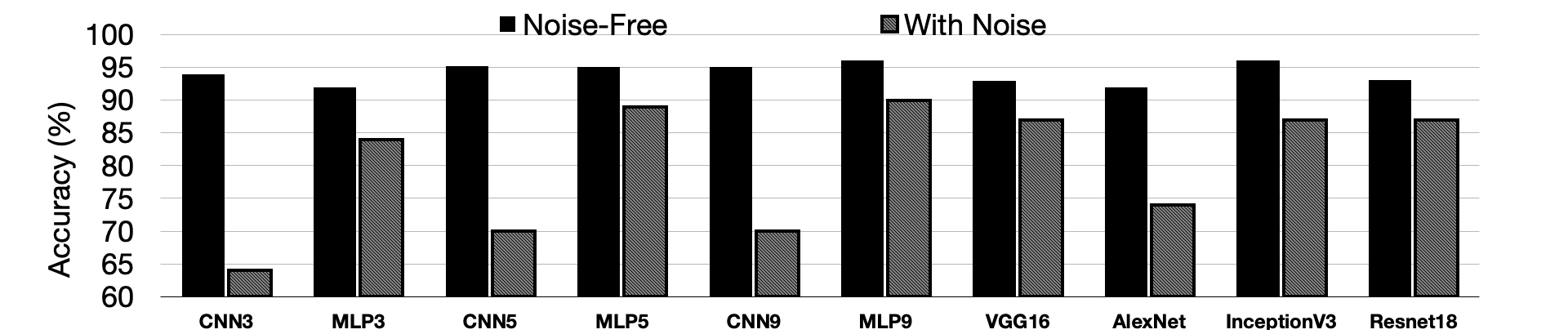}  
  \caption{MZI}
  \label{fig:sub-second}
\end{subfigure}
\newline
\caption{Comparison of the effect of photonic device noise on accuracy using (a) MRR and (b) MZI implementation.}
\label{fig:accuracy_experiment}
\end{figure*}

\begin{figure*}[ht]
    \centering
    \includegraphics[height=0.21\linewidth]{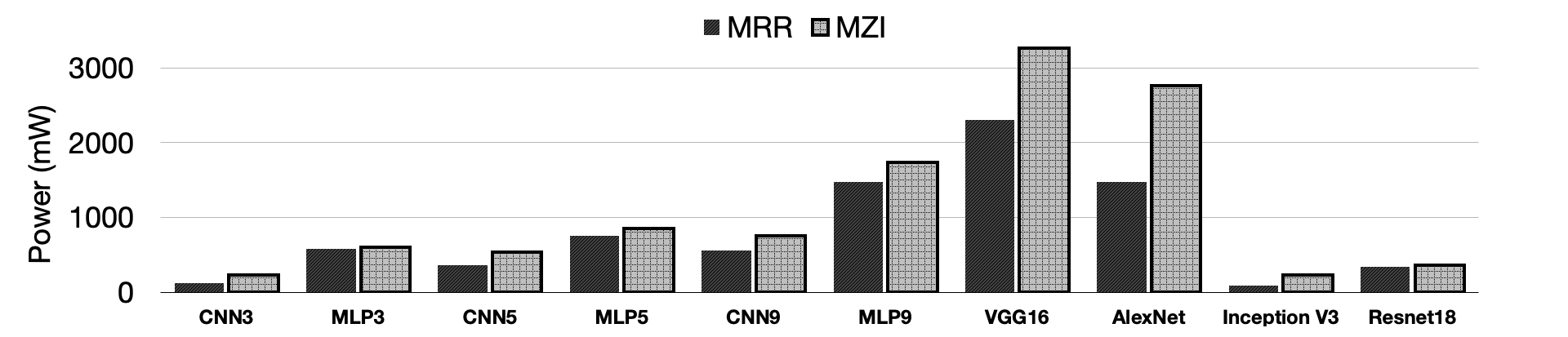}
    \caption{Power estimation of commonly used neural network architectures using photonic components.} 
    \label{fig:power}
\end{figure*}

In this section we present two class of results namely, the functional performance and the power estimation. Figure \ref{fig:accuracy_experiment} represents the comparison of the accuracy of various common neural network architectures for the classification task on the MNIST dataset. The CNN3, CNN5, and CNN9 represent three convolutional neural networks with 3, 5, and 9 convolutional layers and 16 kernels per layer. Similarly MLP3, MLP5, and MLP9 are fully-connected multi-layer perceptron networks. Similarly, \textit{VGG16}, \textit{AlexNet}, \textit{InceptionV3}, and \textit{Resnet} are commonly used deep neural network models \cite{canziani2016analysis}. As we expected the introduction of photonic device noise adversely impacts the accuracy. However, it seems that MRR based implementations suffer less compared to MZI counterparts. In the second experiment we estimated photonic power for the same class of neural network application with both MRR and MZI implementations. Figure \ref{fig:power} summarizes the results. While for most of the architectures the power estimation of MRR-based and MZI-based systems closely follow each other, as the number of network parameters increase, for instance for \textit{VGG16} and \textit{AlexNet} the gap between power consumption of the two device implementations widens.

\section{Conclusion}
 In this paper we proposed a structured methodology and a tool that can be adopted in the design of post-Moore's law accelerators using novel technologies. We considered the case of photonic neuromorphic accelerator design, where  there is a lack of simulation tools that can bridge the design abstraction gap. Rather than building our tool from grounds up, we extended an existing and familiar open-source tool, namely the Google Tensorflow. This allowed us to take advantage of many optimized low-level and mid-level functionalities and kernels, while extending Tensorflow libraries with functional and measurement modules, as well as models to account for photonic device-specific noise sources. We showed that our tool can be used for design space explorations by selecting candidate devices based on their power and functional performance metrics.
\bibliographystyle{IEEEtran}
\bibliography{./references.bib}

\end{document}